\documentclass[a4paper,10pt,eqno]{article}
\usepackage{theorem}
\usepackage{latexsym,amssymb,amsfonts,amsmath}
\usepackage{graphicx}
\newtheorem{thm}{Theorem}[section]
\newtheorem{lem}[thm]{Lemma}

\newtheorem{pro}[thm]{Proposition}

\theoremheaderfont{\scshape}

\newcommand{\ZM}{\mathbb{Z}}

\newcommand{\CM}{\mathbb{C}}

\newcommand{\PM}{\mathbb{P}}

\newcommand{\ket}[1]{|#1\rangle}

\title{{\Large {\bf Localization of an inhomogeneous discrete-time \\ 
quantum walk on the line}}}
\author{
{\small Norio Konno}\\
{\scriptsize Department of Applied Mathematics, 
Faculty of Engineering, 
Yokohama National University}\\
{\scriptsize Hodogaya, Yokohama 240-8501, Japan}\\
{\scriptsize e-mail: konno@ynu.ac.jp}\\
}
\date{\empty }
\begin{document}
\maketitle

\par\noindent
\begin{small}
\par\noindent
{\bf Abstract}. We investigate a space-inhomogeneous discrete-time quantum walk in one dimension. We show that the walk exhibits localization by a path counting method.

\footnote[0]{
{\it Abbr. title:} Localization of an inhomogeneous quantum walk
}
\end{small}

\setcounter{equation}{0}
\section{Introduction}
As a quantum counterpart of the classical random walk, the quantum walk (QW) has recently attracted much attention for various fields. There are two types of QWs. One is the discrete-time walk and the other is the continuous-time one. The discrete-time QW in one dimension (1D) was intensively studied by Ambainis et al. \cite{AmbainisEtAl2001}. One of the most striking properties of the 1D QW is the spreading property of the walker. The standard deviation of the position grows linearly in time, quadratically faster than classical random walk. The review and book on QWs are Kempe \cite{Kempe2003}, Kendon \cite{Kendon2007}, Venegas-Andraca \cite{VAndraca2008}, Konno \cite{Konno2008b}, for examples.

In the present paper we focus on discrete-time case. The model considered here is a space-inhomogeneous two-state 1D QW. The two-state corresponds to left and right chiralities defined in the next section. Let $p_{n} (0)$ denote the probability that the walker returns to the origin at time $n$. The model is said to exhibit localization if $\lim_{n \to \infty} \> p_{2n} (0) >0$. The homogeneous two-state 1D QW except a trivial case does not exhibit localization, see \cite{AmbainisEtAl2001}, for example. The decay order of $p_{2n} (0)$ is closely related to the recurrence. As for the recurrence property of QWs, see \v{S}tefa\v{n}\'ak et al. \cite{StefanakEtAl2008a,StefanakEtAl2008b,StefanakEtAl2009}. Localization of the homogeneous model was shown for a three-sate 1D QW in \cite{InuiEtAl2005}, a four-state 1D QW in \cite{InuiKonno2005}, and a multi-state QW on tree in \cite{ChisakiEtAl2009}. Mackay et al. \cite{MackayEtAl2002} and Tregenna et al. \cite{TregennaEtAl2003} found numerically that a homogeneous 2D QW exhibits localization. Inui et al. \cite{InuiEtAl2004} and Watabe et al. \cite{WatabeEtAl2008} showed the phenomenon. In higher dimensions, a $d$-dimensional homogeneous tensor-product coin model does not exhibit localization \cite{StefanakEtAl2008b}. Oka et al. \cite{OkaEtAl2005} analyzed localization of a two-state QW on a semi-infinite 1D lattice, which is closely related to the Landau-Zener transition dynamics. Through numerical simulations, Buerschaper and Burnett \cite{Buerschaper2004} and W\'ojcik et al. \cite{WojcikEtAl2004} reported that the dynamics of the two-state 1D QWs exhibits from dynamical localization, spreading more slowly than in the classical case, to linear diffusion like the homogeneous two-state 1D QW as the period of the perturbation is varied. Linden and Sharam \cite{LindenSharam2009} investigated a similar inhomogeneous two-state 1D QW where the inhomogeneity is periodic in position. They showed that, depending on the period $2k$, the QW can be bounded for even $k$ and unbounded for odd $k$ in time. The former case corresponds to localization and the latter case to delocalization. An interesting question is whether localization emerges even for a simpler inhomogeneous two-state 1D QW compared with the previous models. The present paper gives an affirmative answer to the question. Our result could be useful for quantum information processing by controlling the spreading of the walker.

The rest of the paper is organized as follows. Section 2 gives the definition of our model. In Sect. 3, we present our main result (Theorem {\rmfamily \ref{thm1}}) of this paper. Section 4 is devoted to the proof of Proposition {\rmfamily \ref{pro0}}. In Sect. 5, we prove Theorem {\rmfamily \ref{thm1}}. Finally, we consider some related models in Sect. 6. In contrast to our model, we show that a similar simple inhomogeneous two-state 1D QW, whose limit theorem presented in \cite{Konno2009}, does not exhibit localization. Moreover, for the corresponding classical random walk also, localization does not emerge.

\section{Definition of the walk}
In this section, we give the definition of the inhomogeneous two-state QW on $\ZM$ considered here, where $\ZM$ is the set of integers. The discrete-time QW is a quantum version of the classical random walk with additional degree of freedom called chirality. The chirality takes values left and right, and it means the direction of the motion of the walker. At each time step, if the walker has the left chirality, it moves one step to the left, and if it has the right chirality, it moves one step to the right. Let define
\begin{eqnarray*}
\ket{L} = 
\left[
\begin{array}{cc}
1 \\
0  
\end{array}
\right],
\qquad
\ket{R} = 
\left[
\begin{array}{cc}
0 \\
1  
\end{array}
\right],
\end{eqnarray*}
where $L$ and $R$ refer to the left and right chirality state, respectively.  

For the general setting, the time evolution of the walk is determined by a sequence of $2 \times 2$ unitary matrices, $\{ U_x : x \in \ZM \}$, where
\begin{align*}
U_x =
\left[
\begin{array}{cc}
a_x & b_x \\
c_x & d_x
\end{array}
\right],
\end{align*}
with $a_x,b_x,c_x,d_x \in \mathbb C$ and $\CM$ is the set of complex numbers.  The subscript $x$ indicates the location. The matrices $U_x$ rotate the chirality before the displacement, which defines the dynamics of the walk. To describe the evolution of our model, we divide $U_x$ into two matrices:
\begin{eqnarray*}
P_x =
\left[
\begin{array}{cc}
a_x & b_x \\
0 & 0 
\end{array}
\right], 
\quad
Q_x=
\left[
\begin{array}{cc}
0 & 0 \\
c_x & d_x 
\end{array}
\right],
\end{eqnarray*}
with $U_x=P_x+Q_x$. The important point is that $P_x$ (resp. $Q_x$) represents that the walker moves to the left (resp. right) at position $x$ at each time step.

For a given sequence $\{ \omega_x : x \in \ZM \}$ with $\omega_x \in [0, 2 \pi)$, our previous paper \cite{Konno2009} treated the following $U_x$: 
\begin{align}
U_x = U_x (\omega_x) = 
\frac{1}{\sqrt{2}}
\left[
\begin{array}{cc}
e^{i \omega_x} & 1 \\
1 & -e^{-i \omega_x} 
\end{array}
\right].
\label{akina}
\end{align}
In the present paper, for a given sequence $\{ \omega_x : x \in \ZM \}$, we consider 
\begin{align}
U_x = U_x (\omega_x) = \frac{1}{\sqrt{2}}
\left[
\begin{array}{cc}
1 & e^{i \omega_x} \\
e^{-i \omega_x} & -1
\end{array}
\right].
\label{seiko}
\end{align} 
In particular, here we concentrate on a simple inhomogeneous model depending only on a one-parameter $\omega \in [0,2\pi)$ as follows: 
\begin{align}
U_0 = U_0 (\omega), \quad U_x = U_x (0) \quad \hbox{if} \> x \not= 0. 
\label{seikou}
\end{align}
So when $\omega \not=0$, our model is homogeneous except the origin. If $\omega = 0$, then this model becomes homogeneous and is equivalent to the {\it Hadamard walk} determined by the Hadamard gate $U_x = U_x (0) \equiv H$:
\begin{eqnarray*}
H=\frac{1}{\sqrt2}
\left[
\begin{array}{cc}
1 & 1 \\
1 &-1 
\end{array}
\right].
\end{eqnarray*}
In this paper, we take $\varphi_{\ast} = {}^T [1/\sqrt{2},i/\sqrt{2}]$ as the initial qubit state, where $T$ is the transposed operator. Then the probability distribution of the Hadamard walk starting from $\varphi_{\ast}$ at the origin is symmetric.

Let $\Xi_{n} (l,m)$ denote the sum of all paths starting from the origin in the trajectory consisting of $l$ steps left and $m$ steps right at time $n$ with $n=l+m$. For example, 
\begin{align*}
\Xi_2 (1,1) &= Q P_0 + P Q_0, \\
\Xi_4 (2,2) &= Q^2 P P_0 + P^2 Q Q_0 + Q P_0 Q P_0 + P Q_0 P Q_0 + P Q_0 Q P_0 + Q P_0 P Q_0. 
\end{align*}
The probability that our quantum walker is in position $x$ at time $n$ starting from the origin with $\varphi_{\ast} (={}^T [1/\sqrt{2},i/\sqrt{2}])$ is defined by 
\begin{align*}
P (X_{n} =x) = || \Xi_{n}(l, m) \varphi_{\ast} ||^2,
\end{align*}
where $n=l+m$ and $x=-l+m$.
The following is the important quantity of this paper.
\begin{align*}
p_n (0) = P (X_{n} =0).
\end{align*}
This is the return probability at time $n$. Remark that $p_{2n+1} (0) = 0$ for $n \ge 0$. For our model with $\omega = \pi$, a direct computation implies
\begin{align*}
p_{2} (0) &= \frac{2}{2^2} = 0.5, \quad
p_{4} (0) = \frac{10}{2^4} = 0.625, \quad 
p_{6} (0) = \frac{40}{2^6} = 0.625, \\
p_{8} (0) &= \frac{170}{2^8} = 0.66406 \ldots, \quad
p_{10} (0) = \frac{680}{2^{10}} = 0.66406 \ldots, \quad 
p_{12} (0) = \frac{2600}{2^{12}} = 0.63476 \ldots.
\end{align*}
In fact, as a consequence of our main result (Theorem {\rmfamily \ref{thm1}}), we have $\lim_{n \to \infty} p_{2n} (0) = (4/5)^2 = 0.64.$ Therefore the QW with $\omega = \pi$ exhibits localization. On the other hand, for the Hadamard walk case (i.e., $\omega =0$), 
\begin{align*}
p_{2}^{(H)} (0) &= \frac{2}{2^2} = 0.5, \quad
p_{4}^{(H)} (0) = \frac{2}{2^4} = 0.125, \quad 
p_{6}^{(H)} (0) = \frac{8}{2^6} = 0.125, \\
p_{8}^{(H)} (0) &= \frac{18}{2^8} = 0.07031 \ldots, \quad
p_{10}^{(H)} (0) = \frac{72}{2^{10}} = 0.07031 \ldots, \quad 
p_{12}^{(H)} (0) = \frac{200}{2^{12}} = 0.04882 \ldots.
\end{align*}
Superscript $(H)$ denotes the Hadamard walk. In this case, it is known that $\lim_{n \to \infty} p_{2n} (0) = 0,$ (see Sect. 6). So the QW with $\omega = 0$ does not exhibit localization.

\section{Our result}
In this section, we present our main result on the inhomogeneous two-state 1D QW. Let 
\begin{align}
\Psi_{2n} (0) =  
\left[
\begin{array}{cc}
\Psi_{2n}^{(L)} (0) \\
\Psi_{2n}^{(R)} (0) 
\end{array}
\right]
= \Xi_{2n}(n, n) \varphi_{\ast}
\end{align}
for $n \ge 0$. This is the probability amplitude at the origin at time $2n$, where the upper (lower) component corresponds to the left (right) chirality. Remark that $\Psi_{2n+1} (0) = {}^T [\Psi_{2n+1}^{(L)} (0), \Psi_{2n+1}^{(R)} (0)] = {}^T [0,0]$ for $n \ge 0$. Let $\PM = \{1,2, \ldots \}$. Then we have
\begin{pro}
\label{pro0} For $n \ge 1$, 
\begin{align*}
\Psi_{2n} (0) 
&= \frac{1}{\sqrt{2}} \> \sum_{k=1}^n \sum_{\scriptstyle (a_1, \ldots , a_k) \in \PM^k : \atop \scriptstyle a_1 + \cdots + a_k =n} \left( \prod_{j=1}^k r^{\ast}_{2 a_j-1} \right) 
\times 
\left[
\begin{array}{cc}
\frac{1- \mu_{+}}{C_{+}^2} \left( \frac{\gamma_{+}}{2} \right)^k 
+ \frac{1- \mu_{-}}{C_{-}^2} \left( \frac{\gamma_{-}}{2} \right)^k \\
- \left\{ \frac{\mu_{+}(1- \mu_{+})}{C_{+}^2} \left( \frac{\gamma_{+}}{2} \right)^k + \frac{\mu_{-}(1- \mu_{-})}{C_{-}^2} \left( \frac{\gamma_{-}}{2} \right)^k \right\} i 
\end{array}
\right],
\end{align*}
where 
\begin{align*}
\gamma_{\pm} 
&= - \cos \omega \pm i \sqrt{1 + \sin^2 \omega}, \quad 
\mu_{\pm} = \sin \omega \mp \sqrt{1 + \sin^2 \omega}, 
\\
C_{\pm} 
&= \sqrt{2 \left\{ (1+\sin^2 \omega) \mp \sin \omega \sqrt{1 + \sin^2 \omega} \right\} },
\\
\sum_{n=1}^{\infty} \> r_{n}^{\ast} z^n 
&= \frac{-1 - z^2 + \sqrt{1 + z^4}}{z}.
\end{align*}
\end{pro}
The proof is given in Sect. 4. By using this proposition, the following main result of this paper can be obtained. As for the proof of the theorem, see Sect. 5.
\begin{thm}
\label{thm1}
For our inhomogeneous two-state 1D QW with the parameter $\omega \in [0, 2 \pi)$ defined by Eqs. \ref{seiko} and \ref{seikou}, we have 
\begin{align*}
\lim_{n \to \infty} \> p_{2n} (0) = \left( \frac{2(1-\cos \omega)}{3 - 2 \cos \omega} \right)^2 =: c (\omega),
\end{align*}
where $p_{2n} (0)$ is the return probability at the origin at time $2n$.
\end{thm}
We present some properties on the above limit $c (\omega)$ (see Fig. 1). (i) $c (\omega) = c (2 \pi - \omega)$. (ii) $c (\omega)$ is strictly increasing in $\omega \in [0, \pi]$. (iii) $c (0) = 0 \le c (\omega) \le c(\pi)=(4/5)^2$ for any $\omega \in [0, \pi]$. Therefore if the model is inhomogeneous, i.e., $\omega \in (0, 2 \pi)$, then it exhibits localization, i.e., $ c (\omega) > 0.$ When $\omega$ is the uniform distribution on $[0, 2 \pi)$, we see that $E[c (\cdot)] = (25 - 7 \sqrt{5})/25 = 0.3739 \ldots$, where $E[c (\cdot)]$ is the expectation of $c (\omega)$. As we will discuss in the last section, for another inhomogeneous two-state 1D QW defined by Eq. \ref{akina}, we have $\lim_{n \to \infty} \> p_{2n} (0) =0$. That is, the QW does not exhibit localization.

\par
\
\par
\begin{figure}[h]
\begin{center}
\includegraphics[width=4cm]{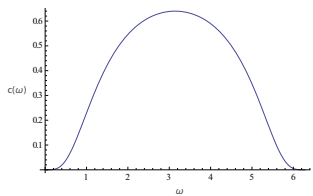}
\end{center}
\caption{The plot of $c(\omega)$}
\label{Figure1}
\end{figure}

\section{Proof of Proposition {\rmfamily \ref{pro0}}}
In this section, we prove Proposition {\rmfamily \ref{pro0}} by using a path counting approach. To do so, we first consider the Hadamard walk starting from location $m \> (\ge 1)$ on $\ZM_{+} = \{ 0, 1, 2, \ldots \}$ with an absorbing boundary at $0$ (see Ambainis et al. \cite{AmbainisEtAl2001} for more details, for example). In this model, $P_0$ and $Q_0$ do not appear, since we consider only $\{ U_x \equiv H : x \ge 1\}$. Therefore the definition of the walk yields  
\begin{align*}
U_x = H = \frac{1}{\sqrt{2}}
\left[
\begin{array}{cc}
1 & 1 \\
1 & -1
\end{array}
\right],
\end{align*}
for $x \ge 1$. Then 
\begin{align*}
P_x = P = \frac{1}{\sqrt{2}}
\left[
\begin{array}{cc}
1 & 1 \\
0 & 0
\end{array}
\right],
\quad
Q_x = Q = \frac{1}{\sqrt{2}}
\left[
\begin{array}{cc}
0 & 0 \\
1 & -1
\end{array}
\right] \qquad (x \ge 1).
\end{align*}
Let $\Xi^{(\infty,m)} _n$ be the sum over possible paths for which the particle first hits 0 at time $n$ starting from $m$. For example, 
\begin{align*} 
\Xi^{(\infty,1)} _5  = P^2 Q P Q + P^3 Q^2.
\end{align*} 
We introduce $R$ and $S$ as follows:
\begin{align*}
R =
\frac{1}{\sqrt{2}}
\left[
\begin{array}{cc}
1 & -1 \\
0 & 0 
\end{array}
\right], 
\quad
S=
\frac{1}{\sqrt{2}}
\left[
\begin{array}{cc}
0 & 0 \\
1 & 1 
\end{array}
\right].
\end{align*} 
We should remark that $P, Q, R$ and $S$ form an orthonormal basis of the vector space of complex $2 \times 2$ matrices with respect to the trace inner product $\langle A | B \rangle = $ tr$(A^{\ast}B)$, where $\ast$ means the adjoint operator. Therefore $\Xi^{(\infty,m)} _n$ can be written as
\begin{align*} 
\Xi^{(\infty,m)} _n = p^{(\infty,m)} _n P + q^{(\infty,m)} _n Q + r^{(\infty,m)} _n R + s^{(\infty,m)} _n S .
\end{align*}
Noting the definition of $\Xi^{(\infty,m)} _n$, we see that for $m \ge 1$, 
\begin{align*} 
\Xi^{(\infty,m)} _n = \Xi^{(\infty,m-1)} _{n-1} P + \Xi^{(\infty,m+1)} _{n-1} Q.\end{align*}
Then we have
\begin{align*}
p^{(\infty,m)} _n 
&= \frac{1}{\sqrt{2}} \> p^{(\infty,m-1)} _{n-1} + \frac{1}{\sqrt{2}} \> r^{(\infty,m-1)} _{n-1}, 
\quad  
q^{(\infty,m)} _n = - \frac{1}{\sqrt{2}} \> q^{(\infty,m+1)} _{n-1} + \frac{1}{\sqrt{2}} \> s^{(\infty,m+1)} _{n-1}, \\ 
r^{(\infty,m)} _n 
&= \frac{1}{\sqrt{2}} \> p^{(\infty,m+1)} _{n-1} - \frac{1}{\sqrt{2}} \> r^{(\infty,m+1)} _{n-1}, 
\quad 
s^{(\infty,m)} _n = \frac{1}{\sqrt{2}} \> q^{(\infty,m-1)} _{n-1} + \frac{1}{\sqrt{2}} \> s^{(\infty,m-1)} _{n-1}. 
\end{align*}
From the definition of $\Xi^{(\infty,m)} _n$, it is easily shown that there exist only two types of paths, that is, $P \ldots P$ and $P \ldots Q$. Therefore we see that $q^{(\infty,m)} _n=s^{(\infty,m)} _n=0$ for $n \ge 1$. We introduce generating functions of $p^{(\infty,m)} _n$ and $r^{(\infty,m)} _n$ as follows:
\begin{align*}
p^{(\infty,m)}  (z) = \sum_{n=1} ^{\infty} p^{(\infty,m)} _n z^n, \quad r^{(\infty,m)}  (z) = \sum_{n=1} ^{\infty} r^{(\infty,m)} _n z^n.
\end{align*}
Then we get
\begin{align*}
p^{(\infty,m)} (z) &= \frac{z}{\sqrt{2}} \> p^{(\infty,m-1)} (z) + \frac{z}{\sqrt{2}} \> r^{(\infty,m-1)} (z), \\
r^{(\infty,m)} (z) &= \frac{z}{\sqrt{2}} \> p^{(\infty,m+1)} (z) - \frac{z}{\sqrt{2}} \> r^{(\infty,m+1)} (z). 
\end{align*} 
Solving these, we see that both $p^{(\infty,m)} (z)$ and $r^{(\infty,m)}  (z)$ satisfy the same recurrence:   
\begin{align*} 
p^{(\infty,m+2)} (z) + \sqrt{2} \> \left( {1 \over z} - z \right) p^{(\infty,m+1)} (z) - p^{(\infty,m)} (z) &= 0,
\\
r^{(\infty,m+2)} (z) + \sqrt{2} \> \left( {1 \over z} - z \right) r^{(\infty,m+1)} (z) - r^{(\infty,m)} (z) &= 0.
\end{align*} 
From the characteristic equations with respect to the above recurrences, we have the same roots: 
\begin{align*}
\lambda_\pm=\frac{-1 + z^2 \pm\sqrt{1+z^4}}{\sqrt{2}z}.
\end{align*}
The definition of $\Xi^{(\infty,1)} _n$ gives $p^{(\infty,1)}_n = 0 \> (n \ge 2)$ and $p_1 ^{(\infty,1)} =1$. So we have $p^{(\infty,1)}  (z) = z$. Moreover noting $\lim_{m \to \infty} p^{(\infty,m)} (z) < \infty$, the following explicit form can be obtained:
\begin{align*}
p^{(\infty,m)} (z) = z \lambda_ +^{m-1}, \quad r^{(\infty,m)} (z) = \frac{-1+\sqrt{1+z^4}}{z} \lambda_+^{m-1}.
\end{align*}
Therefore for $m=1$,
\begin{align*}
r^{(\infty,1)} (z) = \frac{-1+\sqrt{1+ z^4}}{z}.
\end{align*} 
Next we consider the Hadamard walk starting from location $m (\le -1)$ on $\ZM_{-} = \{ 0, -1, -2, \ldots \}$ with an absorbing boundary at $0$. Let $\Xi^{(-\infty,m)} _n$ be the sum over possible paths for which the particle first hits 0 at time $n$ starting from $m \> (\le -1)$. Similarly we see that 
\begin{align*}
q^{(-\infty,m)} (z) = z \lambda_- ^{m+1}, \quad s^{(-\infty,m)} (z) = \frac{1-\sqrt{1+ z^4}}{z} \lambda_-^{m+1}.
\end{align*}
So for $m=-1$, 
\begin{align*}
s^{(- \infty, -1)} (z) = \frac{1 - \sqrt{1 + z^4}}{z}.
\end{align*} 
Remark that $r_{n}^{(\infty,1)} + s_{n}^{(-\infty,-1)} = 0$ for $n \ge 1$. Let $\Xi_n^{+} = \Xi_n^{(\infty, 1)} Q_0$ and $\Xi_n^{-} = \Xi_n^{(-\infty, -1)} P_0$, where
\begin{align*}
P_0 = \frac{1}{\sqrt{2}}
\left[
\begin{array}{cc}
1 & e^{i \omega} \\
0 & 0 
\end{array}
\right], 
\quad
Q_0 = \frac{1}{\sqrt{2}}
\left[
\begin{array}{cc}
0 & 0 \\
e^{-i \omega} & -1
\end{array}
\right].
\end{align*}
That is, $\Xi_n^{+}$ (resp. $\Xi_n^{-}$) is the sum of all paths for which the particle first hits 0 at time $n$ starting from the origin restricted in the region $\ZM_{+}$ (resp. $\ZM_{-}$). Therefore we obtain   
\begin{lem}
\label{lem1}
(i) If $n \ge 4$ and $n$ is even, then
\begin{align*}
\Xi_n^{+} 
= r^{(\infty,1)}_{n-1} \> R Q_0 
= \frac{r^{(\infty,1)}_{n-1}}{2} 
\left[
\begin{array}{cc}
-e^{-i \omega} & 1 \\
0 & 0
\end{array}
\right],
\qquad
\Xi_n^{-} 
= s^{(-\infty,-1)}_{n-1} \> S P_0 
= \frac{s^{(-\infty,-1)}_{n-1}}{2} 
\left[
\begin{array}{cc}
0 & 0 \\
1 & e^{i \omega}
\end{array}
\right],
\end{align*}
where 
\begin{align*}
\sum_{n=1}^{\infty} \> r_{n}^{(\infty,1)} z^n = \frac{-1 + \sqrt{1 + z^4}}{z}, \quad \sum_{n=1}^{\infty} \> s_{n}^{(-\infty,-1)} z^n = \frac{1 - \sqrt{1 + z^4}}{z}.
\end{align*}
(ii) 
\begin{align*}
\Xi_2^{+} 
= P Q_0 
= \frac{-1}{2} 
\left[
\begin{array}{cc}
-e^{-i \omega} & 1 \\
0 & 0
\end{array}
\right], \qquad
\Xi_2^{-} 
= Q P_0 
= \frac{1}{2} 
\left[
\begin{array}{cc}
0 & 0 \\
1 & e^{i \omega}
\end{array}
\right].
\end{align*}
(iii) If $n$ is odd, then 
\begin{align*}
\Xi_n^{+} = \Xi_n^{-} 
= 
\left[
\begin{array}{cc}
0 & 0 \\
0 & 0
\end{array}
\right].
\end{align*}
\end{lem}
Put $\Xi_n^{\ast} = \Xi_n^{+} + \Xi_n^{-}$. From this lemma and $s_{n}^{(-\infty,-1)} = - r_{n}^{(\infty,1)}$for $n \ge 1$, we have
\begin{align*}
\Xi^{\ast}_n
=
\frac{r^{\ast}_{n-1}}{2} \> 
\left[
\begin{array}{cc}
- e^{-i \omega} & 1 \\
-1 & - e^{i \omega}
\end{array}
\right],
\end{align*}
where
\[
r_n^{\ast} = 
\left\{
\begin{array}{cl}
\displaystyle{(-1)^{m-1} \> \frac{(2m-1)!}{2^{2m-1} (m-1)! m!}} & \mbox{if $n=4m-1$ and $m \ge 1$} \\
0 & \mbox{if $n \not= 4m-1, \> n \ge 2$ and $m \ge 1$} \\
-1 & \mbox{if $n =1$}
\end{array}
\right.
\]
In fact, 
\begin{align*}
r_1^{\ast} = -1, \> r_2^{\ast} = 0, \> r_3^{\ast} = 1/2,  \> r_4^{\ast} = r_5^{\ast} = r_6^{\ast} = 0, \> r_7^{\ast} = -1/8, \> r_8^{\ast} = r_9^{\ast} = r_{10}^{\ast} = 0, \ldots.   
\end{align*}
Then the generating function of $r_{n}^{\ast}$ is as follows:
\begin{align*}
\sum_{n=1}^{\infty} \> r_{n}^{\ast} z^n = \frac{-1 - z^2 + \sqrt{1 + z^4}}{z}.
\end{align*}
The definition of $\Xi^{\ast}_{n}$ yields
\begin{align*}
\Psi_{2n} (0) 
= \sum_{k=1}^n \sum_{\scriptstyle (a_1, \ldots , a_k) \in \PM^k : \atop \scriptstyle a_1 + \cdots + a_k =n} \left( \prod_{j=1}^k \Xi^{\ast}_{2 a_j} \right) \> \varphi^{\ast}, 
\end{align*}
where $\PM = \{1,2, \ldots \}.$ Then a little algebra gives
\begin{align*}
\left[
\begin{array}{cc}
- e^{-i \omega} & 1 \\
-1 & - e^{i \omega}
\end{array}
\right]^k \> \frac{1}{\sqrt{2}}
\left[
\begin{array}{cc}
1 \\
i
\end{array}
\right]
= 
\frac{1}{\sqrt{2}} 
\left[
\begin{array}{cc}
\frac{1- \mu_{+}}{C_{+}^2} \gamma_{+}^k 
+ \frac{1- \mu_{-}}{C_{-}^2} \gamma_{-}^k \\
- \left\{ \frac{\mu_{+}(1- \mu_{+})}{C_{+}^2} \gamma_{+}^k + \frac{\mu_{-}(1- \mu_{-})}{C_{-}^2} \gamma_{-}^k \right\} i 
\end{array}
\right],
\end{align*}
where
\begin{align*}
\gamma_{\pm} 
&= - \cos \omega \pm i \sqrt{1 + \sin^2 \omega}, \quad 
\mu_{\pm} = \sin \omega \mp \sqrt{1 + \sin^2 \omega}, 
\\
C_{\pm} 
&= \sqrt{2 \left\{ (1+\sin^2 \omega) \mp \sin \omega \sqrt{1 + \sin^2 \omega} \right\} }.
\end{align*}
Noting that
\begin{align*}
\left( \prod_{j=1}^k \Xi^{\ast}_{2 a_j} \right) \> \varphi^{\ast}
= 
\left( \prod_{j=1}^k r^{\ast}_{2 a_j-1} \right) \> \frac{1}{2^k} \> 
\left[
\begin{array}{cc}
- e^{-i \omega} & 1 \\
-1 & - e^{i \omega}
\end{array}
\right]^k \> 
\frac{1}{\sqrt{2}}
\left[
\begin{array}{cc}
1 \\
i
\end{array}
\right],
\end{align*}
we have the desired conclusion.

\section{Proof of Theorem {\rmfamily \ref{thm1}}}
In this section, we prove our main result, i.e., Theorem {\rmfamily \ref{thm1}}. By Proposition {\rmfamily \ref{pro0}}, we will compute generating function of $\Psi_{n} ^{(L)} (0)$. Put $x_n = r^{\ast}_{2n-1}$ and $u_{\pm}=\gamma_{\pm}/2.$ Then we see that
\begin{align*}
&\sum_{n=1}^{\infty} \Psi_{2n} ^{(L)} (0) z^{2n} 
\\
&
= \frac{1}{\sqrt{2}} \> 
\left[ \frac{1- \mu_{+}}{C_{+}^2} \> \sum_{n=1}^{\infty} \Biggl\{ \sum_{k=1}^n \sum_{\scriptstyle (a_1, \ldots , a_k) \in \PM^k : \atop \scriptstyle a_1 + \cdots + a_k =n} \left( \prod_{j=1}^k x_{a_j} \right) u_+^k \Biggr\} z^{2n} 
\right. \\
&\left.  \qquad \qquad \qquad + \frac{1- \mu_{-}}{C_{-}^2} \> \sum_{n=1}^{\infty} \Biggl\{ \sum_{k=1}^n \sum_{\scriptstyle (a_1, \ldots , a_k) \in \PM^k : \atop \scriptstyle a_1 + \cdots + a_k =n} \left( \prod_{j=1}^k x_{a_j} \right) u_-^k \Biggr\} z^{2n} \right] 
\\
&= \frac{1}{\sqrt{2}} \> 
\left[ \frac{1- \mu_{+}}{C_{+}^2} \> \sum_{k=1}^{\infty} \Biggl\{ \sum_{n=k}^{\infty} \sum_{\scriptstyle (a_1, \ldots , a_k) \in \PM^k : \atop \scriptstyle a_1 + \cdots + a_k =n} \left( \prod_{j=1}^k x_{a_j} \right)  z^{2n} \Biggr\} u_+^k  \right. \\
&\left.  \qquad \qquad \qquad + \frac{1- \mu_{-}}{C_{-}^2} \> \sum_{k=1}^{\infty} \Biggl\{ \sum_{n=k}^{\infty} \sum_{\scriptstyle (a_1, \ldots , a_k) \in \PM^k : \atop \scriptstyle a_1 + \cdots + a_k =n} \left( \prod_{j=1}^k x_{a_j} \right)  z^{2n} \Biggr\} u_-^k  \right] 
\\
&= \frac{1}{\sqrt{2}} \> 
\left[ \frac{1- \mu_{+}}{C_{+}^2} \> \sum_{k=1}^{\infty} \left\{ (-1 - z^2 + \sqrt{1 + z^4}) u_+ \right\}^k + \frac{1- \mu_{-}}{C_{-}^2} \> \sum_{k=1}^{\infty} \left\{ (-1 - z^2 + \sqrt{1 + z^4}) u_- \right\}^k \right]
\\
&= \frac{1}{\sqrt{2}} \> \left\{ \frac{1- \mu_{+}}{C_{+}^2} \> \frac{ (-1 - z^2 + \sqrt{1 + z^4}) u_+}{1 - (-1 - z^2 + \sqrt{1 + z^4}) u_+} + \frac{1- \mu_{-}}{C_{-}^2} \> \frac{ (-1 - z^2 + \sqrt{1 + z^4}) u_-}{1 - (-1 - z^2 + \sqrt{1 + z^4}) u_-} \right\}.
\end{align*}
The first equality comes from Proposition {\rmfamily \ref{pro0}}. As for the third equality, we should remark that for $k=2$, 
\begin{align*}
\sum_{n=2}^{\infty} \sum_{\scriptstyle (a_1, a_2) \in \PM^2 : \atop \scriptstyle a_1 + a_2 =n} x_{a_1} x_{a_2} z^{2n}
&=
\left( \sum_{n=1}^{\infty} r_{2n-1}^{\ast} z^{2n-1} \right)^2 z^2 = \left( \sum_{n=1}^{\infty} r_{n}^{\ast} z^{n} \right)^2 z^2  \\
&= \left( \frac{-1 - z^2 + \sqrt{1 + z^4}}{z} \right)^2 \> z^2 = (-1 - z^2 + \sqrt{1 + z^4})^2.
\end{align*}
In the similar fashion, for general $k \ge 1$, we have 
\begin{align*}
\sum_{n=k}^{\infty} \sum_{\scriptstyle (a_1, \ldots , a_k) \in \PM^k : \atop \scriptstyle a_1 + \cdots + a_k =n} \left( \prod_{j=1}^k x_{a_j} \right)  z^{2n} = (-1 - z^2 + \sqrt{1 + z^4})^k.
\end{align*}
Noting that the initial state $\Psi_{0} ^{(L)} (0) =1/\sqrt{2}$ and  
\begin{align*}
\frac{1- \mu_{+}}{C_{+}^2} + \frac{1- \mu_{-}}{C_{-}^2} = 1, 
\end{align*}
we obtain 
\begin{align*}
\sum_{n=0}^{\infty} \Psi_{n} ^{(L)} (0) z^{n} = \frac{1}{\sqrt{2}} \> \left( \frac{1- \mu_{+}}{C_{+}^2} \> \frac{1}{1 - Z u_+} + \frac{1- \mu_{-}}{C_{-}^2} \> \frac{1}{1 -  Z u_-} \right),
\end{align*}
where $Z = -1 - z^2 + \sqrt{1 + z^4}.$ Next we consider generating function of $\Psi_{n} ^{(R)} (0)$. From the initial state $\Psi_{0} ^{(R)} (0) =i/\sqrt{2}$ and  
\begin{align*}
\frac{\mu_{+}(1- \mu_{+})}{C_{+}^2} + \frac{\mu_{-}(1- \mu_{-})}{C_{-}^2} = -1,
\end{align*}
we similarly get
\begin{align*}
\sum_{n=0}^{\infty} \Psi_{n} ^{(R)} (0) z^{n} = \frac{-i}{\sqrt{2}} \> \left\{ \frac{\mu_+(1- \mu_{+})}{C_{+}^2} \> \frac{1}{1 - Z u_+} + \frac{\mu_-(1- \mu_{-})}{C_{-}^2} \> \frac{1}{1 - Z u_-} \right\}.
\end{align*}
Therefore we have
\begin{align*}
\sum_{n=0}^{\infty} \Psi_n ^{(L,\Re)} (0) z^n 
&= \sum_{n=0}^{\infty} \Psi_n ^{(R,\Im)} (0) z^n 
= \frac{2+ Z \cos \omega}{\sqrt{2} (2+ 2 Z \cos \omega + Z^2)}, 
\\
\sum_{n=0}^{\infty} \Psi_n ^{(L,\Im)} (0) z^n 
&= \frac{(1 + \sin \omega) Z}{\sqrt{2} (2+ 2 Z \cos \omega + Z^2)}, 
\quad 
\sum_{n=0}^{\infty} \Psi_n ^{(R,\Re)} (0) z^n 
= - \frac{(1 - \sin \omega) Z}{\sqrt{2} (2+ 2 Z \cos \omega + Z^2)}, 
\end{align*}
where $\Psi_n ^{(A,\Re)} (0)$ (resp. $\Psi_n ^{(A,\Im)} (0)$) is the real (resp. imaginary) part of $\Psi_n ^{(A)} (0)$ for $A= L, R$. A direct computation gives
\begin{align*}
\sum_{n=0}^{\infty} \Psi_n ^{(L,\Re)} (0) z^n 
&= \sum_{n=0}^{\infty} \Psi_n ^{(R,\Im)} (0) z^n \\
&= \frac{4 - 3 \cos \omega + 2(1- \cos \omega)^2 z^2 + (2 - \cos \omega) z^4 + (2-  \cos \omega) (1 + z^2) \sqrt{1+ z^4}}{2 \sqrt{2} \> \left\{ 3 - 2 \cos \omega + 2(1- \cos \omega)^2 z^2 + (3 - 2 \cos \omega) z^4 \right\}}, \\
\sum_{n=0}^{\infty} \Psi_n ^{(L,\Im)} (0) z^n 
&= - \frac{(1 + \sin \omega) \left\{ 1 + 2(1- \cos \omega) z^2 + z^4 + (-1 + z^2) \sqrt{1+ z^4} \right\}}{2 \sqrt{2} \> \left\{ 3 - 2 \cos \omega + 2(1- \cos \omega)^2 z^2 + (3 - 2 \cos \omega) z^4 \right\}}, \\
\sum_{n=0}^{\infty} \Psi_n ^{(R,\Re)} (0) z^n 
&= \frac{(1 - \sin \omega) \left\{ 1 + 2(1- \cos \omega) z^2 + z^4 + (-1 + z^2) \sqrt{1+ z^4} \right\}}{2 \sqrt{2} \> \left\{ 3 - 2 \cos \omega + 2(1- \cos \omega)^2 z^2 + (3 - 2 \cos \omega) z^4 \right\}}.
\end{align*}
Then we obtain 
\begin{align*}
\Psi_{2n} ^{(L,\Re)} (0)  
&= \Psi_{2n} ^{(R,\Im)} (0)  
\sim \frac{\sqrt{2} (1 - \cos \omega)}{3 - 2 \cos \omega} \> \cos (n \theta_0), \\
\Psi_{2n} ^{(L,\Im)} (0) 
&\sim - \frac{\sqrt{2} (1 - \cos \omega)(1 + \sin \omega)}{(3 - 2 \cos \omega) \sqrt{1 + \sin^2 \omega}} \> \sin (n \theta_0), \\
\Psi_{2n} ^{(R,\Re)} (0)  
&\sim \frac{\sqrt{2} (1 - \cos \omega)(1 - \sin \omega)}{(3 - 2 \cos \omega) \sqrt{1 + \sin^2 \omega}} \> \sin (n \theta_0),
\end{align*}
where $\sin \theta_0 = (2 - \cos \omega) \sqrt{1 + \sin^2 \omega}/(3 \cos \omega -2), \> \cos \theta_0 = -(1- \cos \omega)^2/(3 \cos \omega -2)$ and $f(n) \sim g(n)$ means $f(n)/g(n) \to 1$ as $n \to \infty$. Concerning the above derivation, see pp.264-265 of \cite{Flajolet2009}, for example. The definition of $p_{2n} (0)$ gives
\begin{align*}
p_{2n} (0) = |\Psi_{2n} ^{(L,\Re)} (0)|^2 + |\Psi_{2n} ^{(L,\Im)} (0)|^2 + |\Psi_{2n} ^{(R,\Re)} (0)|^2 + |\Psi_{2n} ^{(R,\Im)} (0)|^2, 
\end{align*}
so the proof of Theorem {\rmfamily \ref{thm1}} is complete.

\section{Discussion}
In the last section, we consider some relations between our model and other related ones. For the Hadamard walk (homogeneous model), a similar argument yields
\begin{align*}
\sum_{n=0}^{\infty} \Psi_n ^{(L,\Re)} (0) z^n 
&= \sum_{n=0}^{\infty} \Psi_n ^{(R,\Im)} (0) z^n 
= \frac{1}{2 \sqrt{2}} \> \left(  1 + \frac{1+z^2}{\sqrt{1+z^4}} \right), \\
\sum_{n=0}^{\infty} \Psi_n ^{(L,\Im)} (0) z^n 
&= - \sum_{n=0}^{\infty} \Psi_n ^{(R,\Re)} (0) z^n 
= - \frac{1}{2 \sqrt{2}} \> \left(  1 + \frac{-1+z^2}{\sqrt{1+z^4}} \right).
\end{align*}
Therefore we have
\begin{align*}
p_{2n} ^{(H)} (0) \sim \frac{1}{\pi n}.
\end{align*}
As for the result, see \cite{AmbainisEtAl2001}, for example. Then $\lim_{n \to \infty} p_{2n}^{(H)} (0) = 0.$ Moreover, using Proposition 4.3 of \cite{Konno2009}, we obtain $p_{2n}^{\ast} (0) = p_{2n}^{(H)} (0)$. Here $p_{2n}^{\ast} (0)$ is the return probability at the origin at time $2n$ for another inhomogeneous model defined by Eq. \ref{akina}, which was studied in \cite{Konno2009}. Therefore the QW has also the same decay order as the homogeneous walk, i.e., Hadamard walk:
\begin{align*}
p_{2n}^{\ast} (0) \sim \frac{1}{\pi n}.
\end{align*}
So $\lim_{n \to \infty} p_{2n}^{\ast} (0) = 0$, then the QW does not exhibit localization. This is in great contrast to our model.

For the inhomogeneous classical random walk starting from the origin on $\ZM$, we similarly get 
\begin{align}
f^{(c)} (z) = \sum_{n=0}^{\infty} p_n ^{(c)} (0) z^n = \left\{ 1 - \left( \frac{p_0}{p} + \frac{q_0}{q} \right) \> \frac{1 - \sqrt{1-4 p q z^2}}{2} \right\}^{-1},
\label{semisemi}
\end{align}
where $p_{n}^{(c)} (0)$ is the return probability at time $n$ for the classical walk. In this model, a walker at location $x$ moves one step to the left with probability $p_x$ and one step to the right with probability $q_x$ where $p_x + q_x =1$ for any $x \in \ZM$ and $p_x = p, \> q_x = q$ for $x \in \ZM \setminus \{0\}.$ From Eq. \ref{semisemi}, we have 
\begin{align*}
p_{2n} ^{(c)} (0) \sim \frac{2}{\frac{p_0}{p} + \frac{q_0}{q}} \> \frac{(4pq)^n}{\sqrt{\pi n}}.
\end{align*}
Therefore $\lim_{n \to \infty} p_{2n}^{(c)} (0) = 0$ and localization does not occur. If $p \not= q$, then $p_{2n}^{(c)} (0)$ decays exponentially. If $p=q$=1/2, 
\begin{align*}
p_{2n} ^{(c)} (0) \sim \frac{1}{\sqrt{\pi n}}.
\end{align*}
The result of the classical walk is also in contrast with that of our model.
\par
\
\par\noindent
{\bf Acknowledgment.} The author thanks T. Machida and E. Segawa for helpful discussions and comments. This work was partially supported by the Grant-in-Aid for Scientific Research (C) of Japan Society for the Promotion of Science (Grant No. 21540118).
\par
\
\par

\begin{small}
\bibliographystyle{jplain}

\begin{thebibliography}{99}


\bibitem{AmbainisEtAl2001} 
Ambainis, A., Bach, E., Nayak, A., Vishwanath, A., Watrous, J.: 
One-dimensional quantum walks. In: Proceedings of the 33rd Annual ACM Symposium on Theory of Computing, pp. 37--49 (2001)

\bibitem{Kempe2003} 
Kempe, J.: 
Quantum random walks - an introductory overview. Contemporary Physics {\bf 44},  307--327 (2003) 


\bibitem{Kendon2007} 
Kendon, V.: 
Decoherence in quantum walks - a review. Math. Struct. in Comp. Sci. {\bf 17}, 1169--1220 (2007)


\bibitem{VAndraca2008} 
Venegas-Andraca, S. E.: 
Quantum Walks for Computer Scientists. Morgan and Claypool (2008)


\bibitem{Konno2008b} 
Konno, N.: 
Quantum Walks. In: Quantum Potential Theory, Franz, U., and Sch\"urmann, M., Eds., Lecture Notes in Mathematics: Vol. 1954, pp. 309--452, Springer-Verlag, Heidelberg (2008)


\bibitem{StefanakEtAl2008a} 
\v{S}tefa\v{n}\'ak, M., Jex, I., Kiss, T.: 
Recurrence and P\'olya number of quantum walks. 
Phys. Rev. Lett. {\bf 100}, 020501 (2008) 


\bibitem{StefanakEtAl2008b} 
\v{S}tefa\v{n}\'ak, M., Kiss, T., Jex, I.:
Recurrence properties of unbiased coined quantum walks on infinite $d$-dimensional lattices.
Phys. Rev. A {\bf 78}, 032306 (2008)


\bibitem{StefanakEtAl2009} 
\v{S}tefa\v{n}\'ak, M., Kiss, T., Jex, I.: 
Recurrence of biased quantum walks on a line. 
New J. Phys. {\bf 11}, 043027 (2009)


\bibitem{InuiEtAl2005} 
Inui, N., Konno, N., Segawa, E.:
One-dimensional three-state quantum walk. 
Phys. Rev. E {\bf 72}, 056112 (2005)


\bibitem{InuiKonno2005} 
Inui, N., Konno, N.:
Localization of multi-state quantum walk in one dimension. 
Physica A {\bf 353}, 133--144 (2005)


\bibitem{ChisakiEtAl2009} 
Chisaki, K., Hamada, M., Konno, N., Segawa, E.: 
Limit theorems for discrete-time quantum walks on trees. 
Interdisciplinary Information Sciences (in press)


\bibitem{MackayEtAl2002} 
Mackay, T. D., Bartlett, S. D., Stephenson, L. T., Sanders, B. C.:  
Quantum walks in higher dimensions. 
J. Phys. A: Math. Gen. {\bf 35}, 2745--2753 (2002)


\bibitem{TregennaEtAl2003} 
Tregenna, B., Flanagan, W., Maile, R., Kendon, V.: 
Controlling discrete quantum walks: coins and initial states. 
New J. Phys. {\bf 5}, 83 (2003)


\bibitem{InuiEtAl2004} 
Inui, N., Konishi, Y., Konno, N.: 
Localization of two-dimensional quantum walks. 
Phys. Rev. A {\bf 69}, 052323 (2004)


\bibitem{WatabeEtAl2008} 
Watabe, K., Kobayashi, N., Katori, M., Konno, N.: 
Limit distributions of two-dimensional quantum walks. 
Phys. Rev. A {\bf 77}, 062331 (2008)


\bibitem{OkaEtAl2005} 
Oka, T., Konno, N., Arita, R., Aoki, H.:  
Breakdown of an electric-field driven system: a mapping to a quantum walk.  
Phys. Rev. Lett. {\bf 94}, 100602 (2005)


\bibitem{Buerschaper2004} 
Buerschaper, O., Burnett, K.: 
Stroboscopic quantum walks.
quant-ph/0406039


\bibitem{WojcikEtAl2004} 
W\'ojcik, A., {\L}uczak, T., Kurzy\'nski, P., Grudka, A., Bednarska, M.: 
Quasiperiodic dynamics of a quantum walk on the line. 
Phys. Rev. Lett. {\bf 93}, 180601 (2004)


\bibitem{LindenSharam2009} 
Linden, N., Sharam, J.:
Inhomogeneous quantum walks. 
arXiv:0906.3692


\bibitem{Konno2009} 
Konno, N.: 
One-dimensional discrete-time quantum walks on random environments. 
Quantum Inf. Proc. {\bf 8}, 387--399 (2009)


\bibitem{Flajolet2009} 
Flajolet, P., Sedgewick, R.: 
Analytic Combinatorics. 
Cambridge University Press (2009)


\end{thebibliography}

\end{small}

\end{document}